\documentclass[letterpaper,12pt]{article}

\usepackage{fullpage}
\usepackage{sectsty}

\begin{document}

\sectionfont{\large}
\subsectionfont{\normalsize}

\begin{center}
\Large\bfseries
Comparison of Black Hole Generators for the LHC
\end{center}

\bigskip

\begin{center}
\small
Douglas M. Gingrich \\

\medskip

\textit{Department of Physics, University of Alberta, Edmonton, AB T6G
2G7 Canada}\\  
\textit{TRIUMF, Vancouver, BC V6T 2A3 Canada}\\ 
{\footnotesize gingrich@ualberta.ca}

\bigskip

\end{center}

\begin{quotation} \noindent
\textbf{Abstract\ } 
We compare Monte Carlo event generators dedicated to simulating the
production and decay of extra-dimensional black holes at the Large
Hadron Collider.     
\end{quotation}

\begin{quotation} \noindent
\end{quotation}

\section{Introduction and General Comparison}

Studies of black hole production and decay at the Large Hadron Collider
(LHC) have been assisted by Monte Carlo event generators. 
In some situations private generators have been written, but have not
been documented in any detail or made readily available. 
Some examples of these types of generators are TRUENIOR, which was used
in the first studies of black holes at the LHC by Dimopoulos and
Landsberg~\cite{Landsberg1,Landsberg2} and an unnamed generator used in
the first study specific to ATLAS by Tanaka \textit{et
al}.~\cite{Tanaka}. 
These generators are made quick and efficient for specific studies by
averaging over some dynamical quantities, rather then generating them
according to probabilistic distributions. 

Two general purpose Monte Carlo generators for simulating black holes at
the LHC have been significantly documented and made available on public 
websites: CHARYBDIS and CATFISH.  
CHARYBDIS\footnote{CHARYBDIS website:
http://www.ippp.dur.ac.uk/montecarlo/leshouches/generators/charybdis/}
\cite{CHARYBDIS,Harris1} is perhaps the most widely used generator and
has resulted in a number of studies (see for example
Ref.~\cite{Harris2}).  
A relatively new generator is CATFISH\footnote{CATFISH website:
http://www.phy.olemiss.edu/GR/catfish} \cite{CATFISH1,CATFISH2} (Collider
grAviTational FIeld Simulator for black Holes). 
With the advent of two generators, it becomes necessary to compare them
and hence enable an educated judgement to be made on which one to use in
a given study. 

This paper compares CHARYBDIS version 1.003 (24 August 2006) with
CATFISH version 1.1 (19 October 2006).
The information on CHARYBDIS comes from reading the documentation and
FORTRAN code, while the information on CATFISH comes from reading the
documentation. 
The source code for CATFISH is not yet readily available.

Both generators are written in FORTRAN and interfaced to the general
purpose Monte Carlo program PYTHIA~\cite{PYTHIA}. 
CHARYBDIS can alternatively be interfaced to
HERWIG~\cite{HERWIG1,HERWIG2}.
In the HERWIG version the initial black hole is placed into the event
record and assigned the PDG code IDHEP = 40 with name 'BlacHole'.
This has not been implemented in the PYTHIA version.
In both generators the interface is defined by the Les Houches
accord~\cite{housches}.
PYTHIA or HERWIG provide the parton evolution and hadronization, as well
as, standard model particle decays.  

The important effects of angular momentum in the production and decay of
black holes in extra dimensions are not accounted for in either generator.

\section{Comparison of Black Hole Production}

Most studies of black hole production in higher dimensions at the LHC
have used the semiclassical black disk approximation for the parton
cross section. 
The factorization approximation is then used to convolute the parton
cross section with the parton density functions (PDFs) in the proton to
obtain a differential cross section depending on the mass of the black
hole.  
This cross section is sampled in a mass range to determine the relative
probability for producing a black hole of a particular mass.

Both generators use the black disk cross section, which depends only on
the horizon radius.
The Schwarzschild radius is chosen for the horizon radius, and depends on
the number of dimensions and the Planck scale. 
The range of the total number of dimensions allowed by CHARYBDIS is 6 to
11, while that allowed by CATFISH is 7 to 11. 
For the Planck scale, three common definitions are available in
CHARYBDIS, while CATFISH has only one definition.
The definition used in CATFISH is the same as the default (MSSDEF=2) in
CHARYBDIS, which is the Dimopoulos and Landsberg~\cite{Landsberg2}
definition.
The definition of the Planck scale is important at LHC energies,
particularly when comparing with experimental results.

In addition to the black disk cross section, CATFISH can produced black
holes inelastically using two gravitational models. 
The gravitational models use the trapped-surface approach, which sets
limits on the minimum energy that gets trapped behind the event
horizon. 
These models also predict values for the form factor.
The model by Yoshino and Nambu~\cite{Yoshino1} determines the apparent
horizon at the instance of collision, while the model by
Yoshino and Rychkov~\cite{Yoshino2} determines the apparent horizon
using the ``optimal slice''. 
CATFISH assumes the remaining energy not forming the black hole is
radiated as gravitons.
This is treaded by giving the proton beam remnants all the lost energy
and not passing them onto PYTHIA.

CHARYBDIS can be used for either proton-proton or proton-antiproton
collisions, and can be easily interfaced to any of the common parton
density functions using the Les Houches accord or the PDFLIB library.
CATFISH simulates proton-proton collisions and only uses the (stable)
CTEQ5 parton density functions for the proton. 
Both generators allow the momentum scale of the parton density functions
to be either the mass of the black hole or the inverse Schwarzchild
radius. 

When using the generators, one specifies a range of back hole masses to
generate. 
In both generators the minimum mass should not be too close to the
Planck scale. 
CATFISH allows the minimum mass to be specified directly or by specifying a
minimum spacetime length.

\section{Comparison of Black Hole Decay}

The decay of a black hole can be view as a four-phase processes:
balding phase, spin-down phase, Hawking evaporation phase, and Planck
phase. 
The balding and spin-down phases are neglected in both generators. 
The trapped-surface models in CATFISH perhaps compensate for these
neglected phases by allowing for lost energy in black hole formation.

The Hawking evaporation, or Schwarzschild, phase is the most well studied.
In both generators, the particles are treated as massless, including the
gauge bosons and heavy quarks.
Baryon number, colour, and electric charge are conserved in the black
hole production and decay in CHARYBDIS.
CATFISH conserves colour but not baryon number.
Electric charge may or may not be conserved in CATFISH depending on the
option chosen.
The black hole is allowed to decay to all standard model particles,
including the Higgs boson.
A few options exist in CHARYBDIS for excluding the Higgs, or the Higgs,
gauge bosons, and top quark.
In CHARYBDIS the graviton is ignored, while in CATFISH it is
included~\cite{Cardoso1,Cardoso2}. 

For the gauge bosons, their longitudinal degrees of freedom coming form
the Higgs mechanism are taken into account slightly differently.
In CHARYBDIS the longitudinal degrees of freedom are considered as
scalars, while in CATFISH they are included in the counting of the
vector bosons. 

We can think of the Hawking evaporation phase as consisting of two
parts: determination of the particle species and assigning energy to the
decay products.
A particle spices is selected randomly with a probability determined by
its number of degrees of freedom and the ratios of emissivities. 
The degrees of freedom take into account polarization, charge, and colour.
In CHARYBDIS the emitted charge is chosen such that the magnitude of the
black hole charge decreases.
This reproduces some of the features of the charge-dependent emission
spectra whilst at the same time making it easier for the event generator
to ensure that charge is conserved for the full decay. 

The energy assignment to the decay particles in the Hawking evaporation
phase has been implemented in each generator differently. 
In CHARYBDIS the particle spices selected by the method described above
is given an energy randomly according to its extra-dimension decay
spectrum.  
A different decay spectrum is used for fermions and vector bosons,
i.e. the spin statistics factor is taken into account.
A Grey-body or a pure black-body spectrum can be used.
Grey-body effects are included without approximations~\cite{Harris3}.
The grey-body factors are spin-dependent and depend on the number of
dimensions.
The choice of energy is made in the rest frame of the black hole before
emission.
The Hawking temperature of the spectrum is either fixed at the beginning
of the decay or updated after each decay. 
If the Hawking temperature is allowed to vary, it is assumed the decay is 
quasi-stationary in the sense that the black hole has time to come into
equilibrium at each new temperature before the next particle is emitted.
If the Hawking temperature is fixed, it is assumed the decay is sudden
in the sense that the back hole spends most of its time near its
original mass and temperature because that is when it evolves the
slowest. 
The energy of the particle given by the spectrum must be constraint to
conserve energy and momentum.
If the decay is not kinematically possible, two options exist: 1) try
again or 2) go directly to the final stage (Planck phase).
Heavy particle production spectra may be unreliable for choices of
parameters for which the initial Hawking temperature is below the rest
mass of the particle being considered.

In CATFISH the initial energy of the black hole is distributed
democratically among all Hawking quanta with a random smearing of $\pm
10$\%. 
If the minimum spacetime length is zero, the decay proceeds according to
the Hawking evaporation theory.
If the minimum spacetime length is non-zero, the decay proceeds according
to modified thermodynamics based on minimal
length.    
The modified thermodynamic model is a result of assuming a generalized
uncertainty principle~\cite{Cavaglia1,Cavaglia2}, which can be motivated
by string theory and non-commutative quantum mechanics.

How the Hawking evaporation phase ends, and the subsequent fate of the
black hole is not know.
The generators handle the termination of the decay process slightly
differently. 
In CHARYBDIS the evaporation phase ends when the black hole mass and/or
the Hawking temperature reaches the Planck scale. 
An option exists in CHARYBDIS to cause the evaporation phase to end when
a particle emission is selected with an energy greater than the black
hole mass. 
In CATFISH the black hole mass signalling the end of the evaporation
phase is a parameter, or the minimum length is used to set this mass.
The minimum energy for black hole production is not the same as the
mass at which the Hawking radiation stops.

After the Hawking evaporation phase, an isotropic $N$-body phase-space
decay is performed, where $N$ is a parameter.
The probability of a particular spices of particle is again given by its
number of degrees of freedom.
In CHARYBDIS $N$ can be between 2 and 5, while in CATFISH it
can be between 2 and 18.
By setting this parameter to zero in CATFISH a black hole remnant is
formed.
An additional mechanism in CATFISH exist to make the remnant charged or
neutral. 
The black hole remnant is not implemented as an actual particle that
can be treated by the parton evolution code, but is rather an amount of
energy-momentum, and possibly charge, which is simply lost. 
In CHARYBDIS an option exist to use the ``boiling remnant'' model, in
which the black hole decay continues below the Planck scale.
In this case, the minimum remnant mass needs to be set (below the Planck
mass) and the temperature may be limited to a maximum value, both specified
by parameters.

\section{Discussion and Recommendations}

Table~\ref{table1} summarises the differences between the CHARYBDIS and
CATFISH generators.
If the two generators treat an aspects of the production or decay in a
similar way, it is not mentioned in the table.

\begin{table}[htb]
\begin{center}
\caption{\label{table1}Summary of differences between CHARYBDIS and
CATFISH.}
\bigskip
\begin{tabular}{|l|l|l|} \hline
Feature           & CHARYBDIS             & CATFISH                 \\ \hline
HERWIG interface  & yes                   & no                      \\
PDFs              & Les Houches or PDFLIB & CTEQ5                   \\
beam particles    & $p$ or $\bar{p}$      & $p$                     \\
total dimensions  & 6--11                 & 7--11                   \\
Planck scale      & 3 definitions         & 1 definition            \\
inelastic effects & no                    & 2 models                \\
form factor       & optional              & if inelastic model      \\
minimum length    & no                    & yes                     \\
temperature       & variable or fixed     & variable                \\
spectrum options  & black-body            & minimum-length modified \\
$N$-body decay    & 2--5                  & 2--18                   \\
remnant           & no                    & charged or neutral      \\
\hline
\end{tabular}
\end{center}
\end{table}

Although the minimum number of extra dimensions is constrained by
experimental bounds, I see no technical reason why either generator
should disallow low values for the number of dimension.
Some caution would be needed in using the generators with 4 dimensions.
The only reason for restricting high values for the number of dimensions
is to allow a simple and efficient coding of the Euler-Gamma function. 

When comparing results for the generators with experiments it is
important to use the same definition of the Planck scale.
It is convenient to use the proper definition in the generator to avoid
having to scale the horizon radius and then propagate this scale to the
cross section and observables being studied. 
On the other hand, the form factor is easily accounted for after event
generation.

I do not believe two models for inelastic production used in CATFISH is
necessary. 
Both only give lower limits on the cross section and are based on a
common set of assumptions and approximations.
I recommend using the ``optimal cut'' model.

Missing energy in CHARYBDIS comes only from the neutrinos.
While in CATFISH missing energy is also possible due to lost energy in
inelastic production, gravitons and a non-detectable black hole remnant,
if selected.
The effect of inelasticity is to shift the cross section to lower mass
but does not give rise to any unique decay signatures.
Gravitons could cause significant missing energy for high dimensions.

CATFISH includes the graviton and thus uses exact field emissivities. 
Although the number of degrees of freedom of the graviton is relatively
low, the graviton emissivity is highly enhanced as the spacetime
dimensionality is increased.
Therefore the black hole may loss a significant fraction of its mass
into the bulk, resulting in missing energy.

In summary, CATFISH includes more options for the production and decay
of black holes.  
Most importantly, it allows for inelastic collisions and gravitons to be
emitted into the bulk.
Many of the more recent results from black hole
production~\cite{Gingrich} and decay studies are not available in
CHARYBDIS.   
Sometimes these features have been added to private versions of
CHARYBDIS (see for example Ref.~\cite{Koch,Alberghi}).
It would be simple to add these features to a new version of CHARYBDIS.

The effects of the different models in the generators on distributions
of observables can be largely gleaned from a knowledge of the
differences in the generators described here.
Reference~\cite{CATFISH1} provides some useful distributions exhibiting
the effects of the different models.

\section*{Acknowledgments}

I would like to thank the authors of the generators, Bryan Webber and
Marco Cavagli{\`a}, for useful comments on the first draft of this paper.


\end{document}